# An Improved UGS Scheduling with QoE Metrics in WiMAX Network

Tarik ANOUARI[1]            Abdelkrim HAQIQ[1, 2]

1 Computer, Networks, Mobility and Modeling laboratory
Department of Mathematics and Computer
FST, Hassan 1st University, Settat, Morocco
2 e-NGN Research group, Africa and Middle East

*Abstract*— WiMAX (Worldwide Interoperability for Microwave Access) technology has emerged in response to the increasing demand for multimedia services in the internet broadband networks. WiMAX standard has defined five different scheduling services to meet the QoS (Quality of Service) requirement of multimedia applications and this paper investigates one specific scheduling service, i.e. UGS scheduling. In parallel, it was observed that in the difference of the traditional quality assessment approaches, nowadays, current researches are centered on the user perception of the quality, the existing scheduling approaches take into account the QoS, mobility and many other parameters, but do not consider the Quality of Experience (QoE). In order to control the packet transmission rate so as to match with the minimum subjective rate requirements of each user and therefore reduce packet loss and delays, an efficient scheduling approach has been proposed in this paper. The solution has been implemented and evaluated in the WiMAX simulation platform developed based on NS-2. Simulation results show that by applying various levels of MOS (Mean Opinion Score) the QoE provided to the users is enhanced in term of jitter, packet loss rate, throughput and delay.

## I. INTRODUCTION

Habitually, the network has been assessed objectively by measuring some parameters to evaluate the network service quality. This evaluation is known as the QoS of the network. The term QoS refers to the guarantees on the ability of a network to deliver predictable results and a more deterministic performance, so data can be transferred with a minimum delay, packet loss, jitter and maximum throughput. The QoS does not take into account the user's perception of the quality. Another approach which takes into account the user's perception is named QoE, it's the overall acceptability of an application or service, as perceived subjectively by the end user, it groups together user perception, expectations, and experience of application and network performance.

In order to get a more comprehensive view of the quality perceived by end users, QoE it has become increasingly a very interesting area of research. Many related works was presented on analyzing and improving QoE [12] in WiMAX network. The study presented in [14] suggested an estimation method of QoE metrics based on QoS metrics in WiMAX network. The QoE was estimated by using a Multilayer Artificial Neural Network (ANN). The results show an efficient estimation of metrics of QoE with respect to QoS parameters.

In [6, 7, 8], the authors focus on the ANN method to adjust the input network parameters to get the optimum output to satisfy the end users. Especially, the success of the ANN approach depends on the model's capacity to completely learn the nonlinear interactions between QoE and QoS. In [16], Muntean proposes a learner QoE model that considers delivery performance-based content personalization in order to improve user experience when interacting with an online learning system. Simulation results show significant improvements in terms of learning achievement, learning performance, learner navigation and user QoE.

In [3], our study was focused on studying and analyzing QoS performances of VoIP traffic using different service classes in term of throughput, jitter and delay. The simulation results show that UGS service class is the best suited to handle VoIP traffic. This paper proposes a QoE-based model in order to provide best performances in WiMAX network especially for the real-time traffic. The target of this improvement is to schedule traffic of UGS service class.

The rest of this paper is organized as follows. A short description of WiMAX technology is given in section 2. In section 3, a QoE overview background is presented. The proposed QoE model is described in detail in section 4. Simulation environment and performance parameters are presented in section 5. Section 6 shows simulation results and analysis. Finally, section 7 concludes the paper.

## II. WIMAX TECHNOLOGY

WiMAX is a wireless communication standard based on the 802.16 standards [10, 11], the main objective of WiMAX is to provide an Internet broadband connection to a coverage area with a radius of several kilometers. Unlike ADSL (Asymmetric Digital Subscriber Line) or other wired technologies, WiMAX uses radio waves, similar to those used for mobile phone.

WiMAX can be used in point-to-multipoint (PMP) mode in which serving multiple client terminals is ensured from a central base station, and in point-to-point (P2P) mode, in which there is a direct link between the central base station and the subscriber.

PMP mode is less expensive to implement and operate while P2P mode can provide greater bandwidth.

### A. QoS in WiMAX Network

Since QoS support is an important part of WiMAX network, the concept of QoS was introduced natively in WiMAX [18], so this protocol ensures the good operation of a service. Some services are very demanding; VoIP cannot tolerate delay in the transmission of data. WiMAX uses service classes to allow different QoS between each communication.

The concept of QoS mainly depends on the service provided, its sensitivity to transmission errors, its requirement of response time... etc. For VoIP traffic, one of the challenges is related to network congestion and latency, we will need a real-time traffic transfer, with very low latency and low jitter. A complete definition of QoS often refers to the mode of transport of information, although the solution adopted by the network to provide the service must remain transparent to the user.

Satisfying QoS requirement becomes very imperative in IEEE802.16 systems to provide best performance, in particular in the presence of various types of connections, namely the current calls, new calls and the handoff connection.

### B. WiMAX Network Architecture

The architecture of WiMAX network consists of base station named BTS (Base Transceiver Station) or BS (Base Station) and mobile clients or stations (SS Subscriber Station). The base station acts as a central antenna responsible for communicating and serve mobile stations, in their turn, serve clients using WIFI or ADSL. The BS can provide various levels of QoS over its queuing, scheduling, control, signaling mechanisms, classification and routing. Figure 1 shows the architecture of WiMAX network [10, 11].

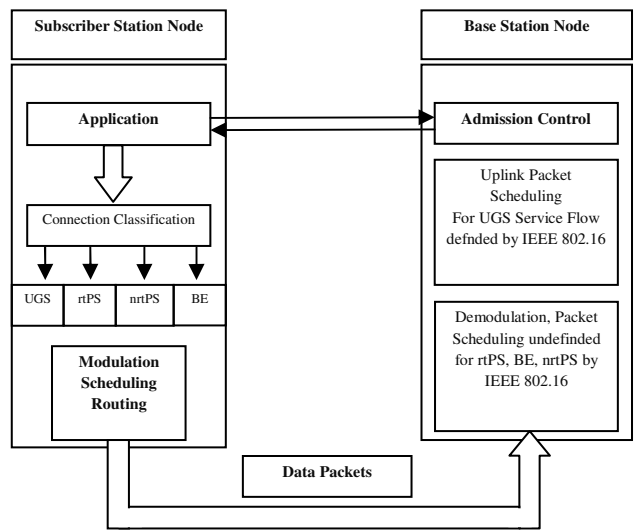

Figure 1: WiMAX Network Architecture

### C. Different Service Classes in WiMAX

Multiple kinds of traffic are considered in WiMAX. QoS is negotiated at the service flow, especially at the establishment of the connection. A modulation and coding technique are set up. To satisfy different types of applications, WiMAX standard has defined four service classes of quality, namely Unsolicited Grant Service (UGS), Best Effort (BE), real-time Polling Service (rtPS) and non-real time Polling Service (nrtPS). The amendment to the IEEE 802.16e standard (802.16e 2005) [1] on mobility includes a fifth type of service class, the extended real-time Polling Service (ertPS). This service is placed between the UGS service and rtPS service. It can serve real-time applications that generate periodic packets of variable size, the example given in the standard is that of a VoIP application with silence suppression.

Some services like VoIP are very demanding in term of QoS, it cannot tolerate delay in data transmission while others have fewer requirements.

Table 1 classifies different service classes of WiMAX and gives their description and QoS parameters.

TABLE I. SERVICE CLASSES IN WiMAX

| Service | Description | QoS parameters |
|---------|-------------|----------------|
| UGS | Real-time data streams comprising fixed size data packets at periodic intervals | Maximum Sustained Rate Maximum Latency Tolerance Jitter Tolerance |
| rtPS | support real-time service flows that periodically generate variable-size data packets | Traffic priority Maximum latency tolerance Maximum reserved rate |
| ertPS | Real-time service flows that generate variable-sized data packets on a periodic basis. | Minimum Reserved Rate Maximum Sustained Rate Maximum Latency Tolerance Jitter Tolerance Traffic Priority |
| nrtPS | Support for non-real-time services that require variable size data grants on a regular basis | Traffic priority Maximum reserved rate Maximum sustained rate |
| BE | Data streams for which no data minimum service level is required. | Maximum Sustained Rate Traffic Priority |

## III. QUALITY OF EXPERIENCE

Quality of Experience (QoE, user Quality of Experience or simply QX) is a subjective measure that reflects the user satisfaction with the service provided (web browsing, phone call, TV broadcast, call to a Call Center).

Today, assessing the quality of experience has become essential for service providers and content providers.

### A. Quality of Experience vs Quality of Service assessment

QoS appeared in the 90 years to describe the quality of the network. Since that time the acronym QoS has been usually used to describe the improved performance realized by hardware and / or software. But with the rapid improvement of Media services, this measure has shown its limitations and many efforts have been made to develop a new metric that reflects more accurately the quality of service provided. This measure is called the QoE.

QoE is a subjective measure of a customer's experiences with a service according to his perception. Indeed, the notion of user experience has been introduced for the first time by Dr. Donald Norman, citing the importance of designing a user centered service [17].

Gulliver and Ghinea [9] classify QoE into three components: assimilation, judgment and satisfaction. The assimilation is a quality assessment of the clarity of the contents by an informative point of view. The judgment of quality reflects the quality of presentation. Satisfaction indicates the degree of overall assessment of the user.

QoE and QoS have become complementary concepts: QoS indicators are used to identify and analyze the causes of network congestions while QoE indicators are used to monitor the quality offered to users. These two solutions used in parallel are a complete system monitoring.

### B. QoE measurement approaches

Two main quality evaluation methodologies are defined, namely objective and subjective performance evaluation. Subjective assessments are carried out by end users who are asked to evaluate the overall perceived quality of the service provided, the most frequently used measurement is the MOS recommended by the International Telecommunication Union (ITU) [13], and it's defined as a numeric value evaluation from 1 to 5 (i.e. poor to excellent).

Objective methods are centered on algorithms, mathematical and/or comparative techniques that generate a quantitative measure of the service provided.

Peter and Bjørn [5] classified the existing approaches of measuring network service quality from a user perception into three classifications, namely: Testing User-perceived QoS (TUQ), Surveying Subjective QoE (SSQ) and Modeling Media Quality (MMQ). The first two approaches collect subjective information from users, whereas the third approach is based on objective technical assessment. Figure 2 [2] gives an overview of the classification of the existing approaches.

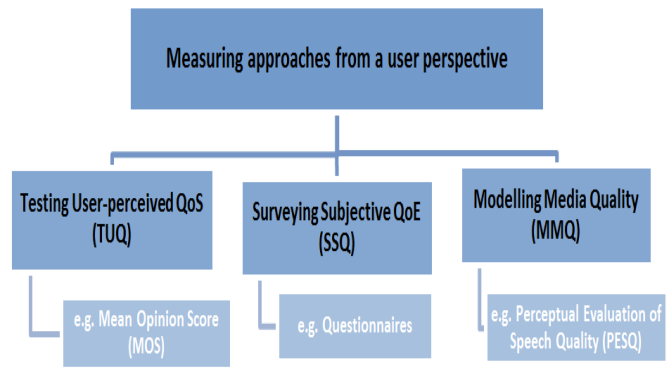

Figure 2. The approaches for measuring network service quality from a user perception

## IV. QoE-BASED SCHEDULING ALGORITHM MODEL

In this section, we propose a QoE-based scheduling approach in WiMAX network, because it's observed that the existing scheduling algorithms take into account QoS but not user perception of the service provided, where every user has different subjective requirement of the system.

### A. Proposed QoE model

In the proposed QoE-based model three QoE levels are used, each user has an initial maximum transmission rate, a minimum subjective rate requirement and a subjective threshold value. The traffic starts with a maximum transmission rate on each user. When the packet loss rate is greater than the user selected threshold (which is chosen at the beginning of the simulation), then each user checks if the transmission rate is higher than the minimum subjective requirement, if yes the transmission rate is decreased, otherwise it's remained at the same level.

In the other hand, if the packet loss rate is less than the selected threshold, then the user checks if the transmission rate is lower than the minimum subjective requirement, if yes the transmission rate is increased, otherwise it's remained at the same level.

The threshold can be selected by the user as a percentage of the data transmission rate, for example, if the user introduces a value of 50 as a threshold then the threshold for packet loss rate is 50%. Figure 3 shows the activity diagram of the proposed model.

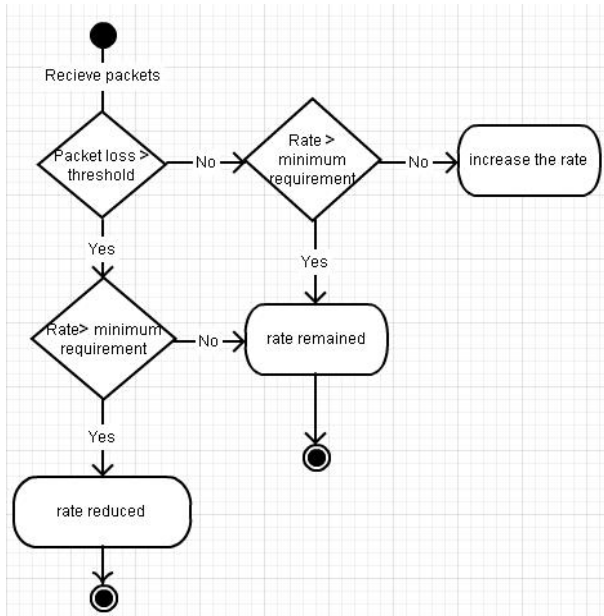

Figure 3: Activity diagram of the proposed QoE-Model

## V. SIMULATION ENVIRONNEMENT

### A. Simulation Model

In this paper, we evaluate the performances of the proposed QoE-based scheduling algorithm, as we consider the Wireless-OFDM PHY layer, our QoE-model is evaluated and compared with the popular WiMAX module developed by NIST (National Institute for Standards and Technologies), which is based on the IEEE 802.16 standard (802.16-2004) and the mobility extension (80216e-2005) [19]. Our simulation scenario consists of creating five wireless users connected to a base station (BS). A sink node is created and attached to the base station to accept packets. A traffic agent is created and then attached to the source node. The Network Simulator (NS-2) [15] is used.

Finally, we set the traffic that produces each node. The first node has run with CBR (Constant Bit Rate) packet size of 200 bytes and interval of "0,0015", the second node has run with CBR packet size of 200 bytes and interval of "0,001", the third node has run with CBR packet size of 200 bytes and interval of "0,001", the fourth node has run with CBR packet size of 200 bytes and interval of "0,001" and fifth node has run with CBR packet size of 200 bytes and interval of "0,0015". The initial transmission rate that produces each node is about "133,3 Kbps", "200 Kbps", "200 Kbps", "200 Kbps" and "133,3 Kbps" respectively. All nodes have the same priority.

Each user has a minimum requirement, so the first user requires minimal traffic rate of "120 Kbps", the second "150 Kbps", the third "150 Kbps", the fourth "150 Kbps" and the fifth "120 Kbps".

The following table summarizes the above description about the produced and required traffic rate of each user.

TABLE II. USER'S TRAFFIC PARAMETERS

| Traffic rate Users | Initial traffic rate (Kbps) | User minimum requirement (Kbps) |
|---|---|---|
| User 1 | 133,33 (200byte/0. 0015) | 120 |
| User 2 | 200 (200byte/0. 001) | 150 |
| User 3 | 200 (200byte/0. 001) | 150 |
| User 4 | 200 (200byte/0. 001) | 150 |
| User 5 | 133.33 (200byte/0. 0015) | 120 |

We use five different thresholds 10%, 20%, 30%, 40% and 50%.

We have used the QoS-included WiMAX module [4] within NS-2.29. This module is based on the NIST implementation of WiMAX [19], it includes the QoS classes as well as the management of the QoS requirements, unicast and contention request opportunities mechanisms, and scheduling algorithms for the UGS, rtPS and BE QoS classes.

The resulted trace files are interpreted and analyzed based on a PERL script, which is an interpretation script software used to extract data from trace files to get throughput, packet loss rate, jitter and delay. The extracted results are plotted in graphs using EXCEL software.

### B. Simulation Parameters

The same simulation parameters are used for both NIST and QOE-based scheduling algorithms, table 3 summarizes the simulation parameters:

TABLE III. SIMULATION PARAMETERS

| Parameter | Value |
|---|---|
| Network interface type | Phy/WirelessPhy/OFDM |
| Propagation model | Propagation/OFDM |
| MAC type | Mac/802.16/BS |
| Antenna model | Antenna/OmniAntenna |
| Service class | UGS |
| packet size | 200 bytes |
| Frequency bandwidth | 5 MHz |
| Receive Power Threshold | 2,025e-12 |
| Carrier Sense Power Threshold | 0,9 * Receive Power Threshold |

| Channel | 3,486e+9 |
|---|---|
| Simulation time | 200s |

### C. Performance Parameters

Main QoS parameters were analyzed in our simulation, namely average throughput, packet loss rate, average jitter and average delay.

## VI. SIMULATION RESULTS AND ANALYSIS

We have performed various simulation scenarios in order to analyse and compare the proposed QoE-based scheduler with the NIST scheduler in term of average throughput, packet loss rate, average delay and average jitter in WiMAX network using UGS service class.

In figure 5, we note that the average throughput in the case of the QoE-based scheduler algorithm is lower than for the NIST scheduler for all flows, whereas the third flow has the largest range between maximum and minimum values.

For the flows 2 and 4 the throughput values are similar for both NIST scheduler and QoE-based scheduler, especially when the QoE threshold is 50%.

The scheduler that takes into account the QoE varied the throughput for different users so as to match with the minimum subjective rate requirements of each user in order to reduce jitter, delays and packet loss.

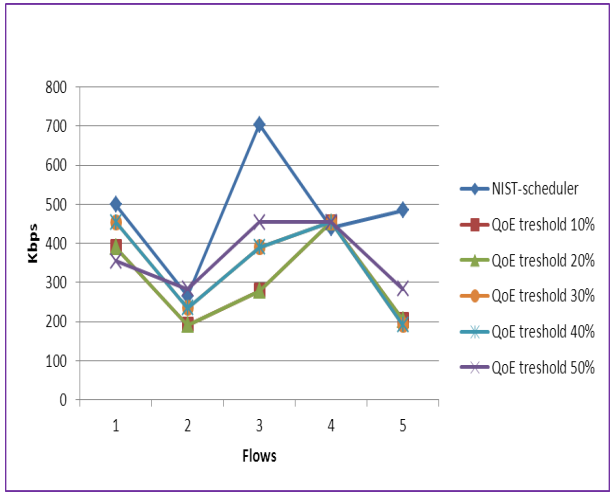

Figure 5. Average throughput

The improvement is noticeable as shown in Figure 6 when the QoE-based scheduler is used. The packet loss rate for all users is reduced while the packet loss rate is similar for both schedulers in the case of flows 3 and 5. The NIST scheduler gives lower performances compared with the QoE-based scheduler in term of packet loss rate.

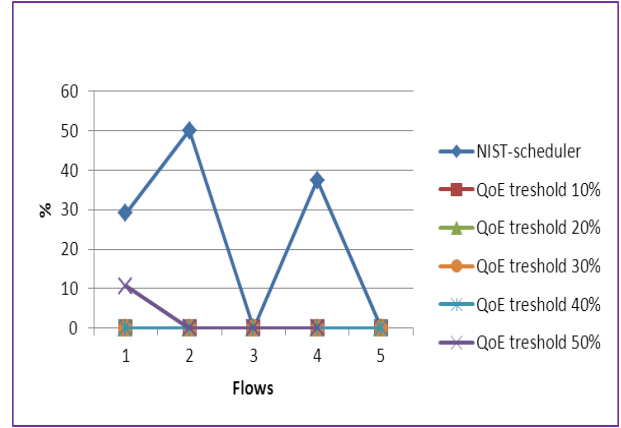

Figure 6. Packet loss rate

It can be observed from the figure 7 that the proposed QoE-based scheduler algorithm has lowest values of average jitter compared with the NIST scheduler by applying different threshold levels, especially for the flows 1, 2 and 3. Average jitter values are identical for flows 4 and 5 for all the threshold levels.

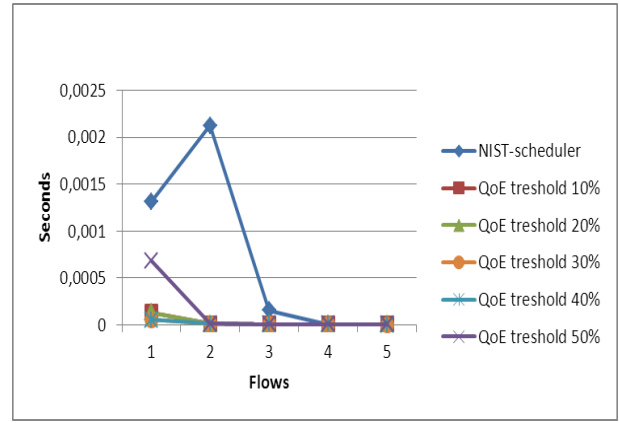

Figure 7. Average Jitter

As shown in figure 8, the QoE-based scheduler outperforms the NIST scheduler, the average transmission packets delay values still lowest in the case of QoE-scheduler, while the two schedulers have similar values for flows 4 and 5.

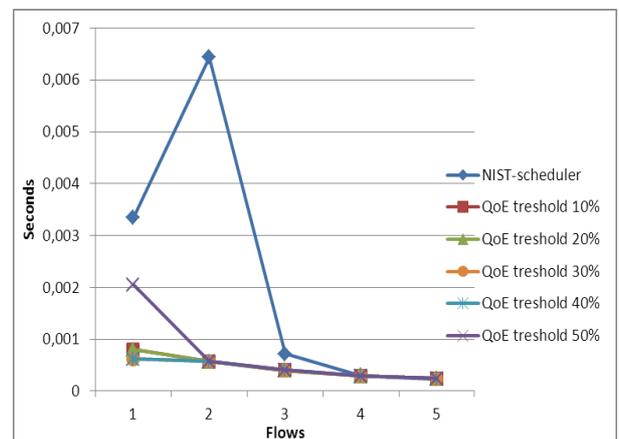

Figure 8. Average Delay

## VII.  CONCLUSION

In this paper, we have proposed a new QoE-based scheduler in order to manage the packet transmission rate for users in WiMAX network. When the packet loss rate exceeds some threshold, there are two cases, either the transmission packet rate is less than the minimum subjective rate requirement, then the user continue to transmit with the same packet transmission rate, otherwise he should reduce it.

The simulations carried out show that the use of different levels of MOS improves the QoE provided to users of WiMAX network. The proposed QoE-model significantly reduced packet loss, delay and jitter, the transmission rate is reduced for each connection, until matching with its minimum subjective rate requirement.

As a future work we may extend this study by adding other parameters like mobility models.